\documentclass[twocolumn,balance,showpacs,preprintnumbers,amsmath,amssymb,superscriptaddress]{revtex4-1}
\usepackage{balance}
\usepackage{amsfonts}
\usepackage{amssymb,epsf}
\usepackage{latexsym}

\usepackage{epsfig}
\begin{document}

\title{\textbf\textrm{{Thermodynamics of a morphological transition in a relativistic gas}}}
\author{Afshin Montakhab}
\email{montakhab@shirazu.ac.ir}
\affiliation{Department of Physics, College of Sciences, Shiraz University, Shiraz 71454, Iran}

\author{Leila Shahsavar}
\affiliation{Department of Physics, College of Sciences, Shiraz University, Shiraz 71454, Iran}

\author{Malihe Ghodrat}
\affiliation{School of Physics, Institute for Research in Fundamental Sciences (IPM), Tehran 19395-5531, Iran}

\date{\today}

\begin{abstract}
Recently, a morphological transition in the velocity distribution of a relativistic gas has been
pointed out which shows hallmarks of a critical phenomenon.  Here, we provide a general framework which
allows for a thermodynamic approach to such a critical phenomenon.  We therefore construct a thermodynamic
potential which upon expansion leads to Landau-like (mean-field) theory of phase transition.  We are
therefore able to calculate critical exponents and explain the spontaneous emergence of ``order parameter" as
a result of relativistic constraints. Numerical solutions which confirm our thermodynamic
approach are also provided. Our approach provides a general understanding of such a transition as well as leading to some new results. Finally, we briefly discuss some possible physical consequences of our results as well as considering the case of quantum relativistic gases.
\vspace{1pc}


\end{abstract}


\maketitle

\section{Introduction}

As early as 1911, F. J\"{u}ttner  provided a relativistic generalization of the famous Maxwell-Boltzmann (MB) velocity distribution whose Gaussian form does not respect the maximal velocity of light, c. He used entropy maximization principle under relativistic energy-momentum conservation constraints to arrive at what is now known as the J\"{u}ttner distribution \cite{Juttner}. For small enough temperatures where the typical velocities are small, the J\"{u}ttner distribution reduces to the classical MB distribution while for high temperatures the relativistic constraints forces it to take on a very different  form. Although the validity of the J\"{u}ttner distribution has been a source of some controversies \cite{Dieckmann, Hees, schieve, weber, Dunkl,CD.EPL}, it has been recently established as the correct relativistic velocity distribution \cite{cubero,ghodrat,montakhab,peano,dunkel,cpc,pre}. However, one important question still remains: what is the temperature scale that determines whether a gas is relativistic or classical? A simple answer may be $\theta\equiv {mc^{2}}/{k_{B}T}\approx1$ where clearly $\theta\gg1 $ is the classical and $\theta\ll1$ the ultra-relativistic limits. Recently, it has been suggested that the J\"{u}ttner distribution exhibits a morphological transition at $\theta_{c}={d+2}$, where d is the spatial dimension of the relativistic gas \cite{mendoza}. That is, for $\theta>\theta_{c}$ the distribution function has a classical form while for $\theta\leq\theta_{c}$ it starts to exhibit an increasingly different form from the classical limit. While this transition temperature is much lower than naive expectation, the more interesting result was that such a transition exhibits some similarity to thermodynamic phase transitions with corresponding (d-independent) critical exponents \cite{mendoza}. However, no reasoning was given as to the origin of such a critical phenomenon. For example, why are the exponents d-independent? What is the physical significance of the order parameter?  Is there a singular behavior (a true hallmark of criticality) as in a ``generalized susceptibility''? Is there a symmetry breaking principle which causes such a phase transition?  In the present work, we provide a general framework which connects the J\"{u}ttner distribution with the thermodynamic theory of phase transitions as described by the Landau theory. In doing so, not only we provide a general framework, we also provide simple answers to the above questions. We also note that such a transition and its physical properties may have important consequences in real world systems for  which $\theta \lesssim 1$, which is not necessarily a very high temperature, for example in graphene as clearly explained in \cite{mendoza}. However, relativistic astrophysics \cite{Wei71,And07} and high energy physics (e.g. quark-gluon plasma \cite{Xu05,Mur10}) are also two important areas of active research where such results may have important consequences. In the following we use natural units, $c=k_{B}=1$, and set $m=1$ without loss of generality.

\section{Results}

Our starting point is a simple observation that in statistical mechanics, equilibrium distributions are related to the thermodynamic potentials via the relation $f=e^{-F}$, thus giving $F=-\ln f$, e.g. $f=\frac{1}{\Omega}$ in the entropy representation where all $\Omega$ micro-states are equally likely \cite{callen}. We therefore simply build such a generalized thermodynamic function using the J\"{u}ttner distribution and use it to calculate various thermodynamic relations. The J\"{u}ttner distribution is given by \cite{ghodrat,montakhab}:
\begin{equation}
f(\vec{v},\vec{u},T)=\frac{A\gamma^{d+2}(v)}{\gamma(u)\big[\exp\big(\frac{1-\vec{u}.\vec{v}}{T}\gamma(u)\gamma(v)-\frac{\mu}{T}\big)+\lambda \big]},
\end{equation}
where $\gamma(v)=(1-v^{2})^{-\frac{1}{2}}$ is the Lorentz factor and $u=|\vec{u}|$ is the average velocity, i.e.,  $\vec{u}=<\vec{v}>$ which is taken to be zero in the co-moving frame, and A is a normalization constant. $\mu$ is the chemical potential and $\lambda=+1, -1, 0$ distinguishes the Fermi, Bose and Boltzmann statistics. For simplicity we set $\mu=\lambda=0$ here. This leads to:
\begin{eqnarray}
F(\vec{v},\vec{u},T)&=&-\ln f =-\ln \big[A \gamma^{-1}(u)\big]-(d+2)\ln \gamma(v) \nonumber\\
&+&\big[\frac{1-\vec{u}.\vec{v}}{T}\gamma(v)\gamma(u)\big].\
\end{eqnarray}
Noting that $u,v\leq1$ and that typically $u\ll1$, we Taylor expand the above expression thus obtaining,
\begin{eqnarray}
F(\vec{v},\vec{u},T)&=&\big[-\ln A+\frac{1}{T}\big]+\frac{1}{2}\big[\frac{1}{T}-(d+2)\big]v^{2} \nonumber\\
&+&\frac{3}{8}\big[\frac{1}{T}-\frac{2}{3}(d+2)\big]v^{4}-\frac{\vec{u}.\vec{v}}{T} +\ldots  \
\end{eqnarray}
Clearly, this has the same form as the Landau functional:
\begin{equation}
G(\phi,h,T)=g(h,T)+a(T)\phi^{2}+b(T)\phi^{4}-h\phi,
\end{equation}
which describes the mean-field theory of critical phase transition at a temperature given by $a(T_{c})=0$, in the absence of ``conjugate field'', $h$. This immediately gives the result $T_{c}=1/(d+2)$, consistent with the previous study \cite{mendoza}. Figure 1 shows the J\"{u}ttner function and the corresponding thermodynamic function $F=-\ln f$ for various temperatures for $d=1$. Note that the morphological transition corresponds to the appearance of new stable minima in the thermodynamic function.  Accordingly, thermodynamic properties are obtained by entropy principle which extremizes the thermodynamic potential, leading to
\begin{equation}
\delta F=0=(\frac{1}{T}-\frac{1}{T_{c}})v_{mp}+(\frac{3}{2T}-\frac{1}{T_{c}})v_{mp}^{3}-\frac{u}{T}+\ldots
\end{equation}
which (for $u=0$) gives:
\begin{eqnarray}
       v_{mp}&=&\left\{ \begin{array}{ccc}
                          0 & ; & t<0 \\
                          \pm\sqrt{2t} & ; & \frac{1}{2}>t>0
                        \end{array}
\right.
 \end{eqnarray}
where $v_{mp}$ is the most probable velocity and
$t\equiv(T-T_{c})/T_{c}$, and the upper limit on $t$ is due to the constraint that $v\leq1$.

Comparing the above with the general Landau theory of phase transition, one immediately realizes that $v_{mp}$ is the order parameter associated with a continuous (second-order) phase transition which occurs at $T=T_{c}=1/(d+2)$ in the absence of the conjugate field $u$.  Several thermodynamic relations follow immediately \cite{goldenfeld}:
\begin{equation}
v_{mp}\sim (T-T_{c})^{\frac{1}{2}}       \   \  ; \  \  (T\gtrsim T_{c})
\end{equation}
\begin{equation}
v_{mp}\sim  u^{\frac{1}{3}}                        \  \   ; \  \   (T=T_{c})
\end{equation}
\begin{equation}
\chi \sim \lim_{u\rightarrow 0} (\frac{\partial v_{mp}}{\partial u})_{T}\sim t^{-1} \  \ ; \  \ (|t|\ll1 )
\end{equation}
leading to $\beta=\frac{1}{2},  \delta=3, \gamma=1$ consistent with mean-field values of critical exponents in the theory of phase transition. Eq.(7) is the same as Eq.(6), while Eq.(8) and (9) follow by simple algebra from Eq.(5).
The critical temperature $T_{c}$ is dependent on physical dimension $d$, however, the critical exponents are independent of $d$. This is because of the mean field-like behavior of our formalism. Mean field theories correspond to high dimensional behavior where irrelevance of fluctuations lead to d-independent critical exponents, while this upper critical dimension is determined by the Ginsburg criterion \cite{kardar}.

\begin{figure}
\centering
\begin{tabular}{cc}
\epsfig{file=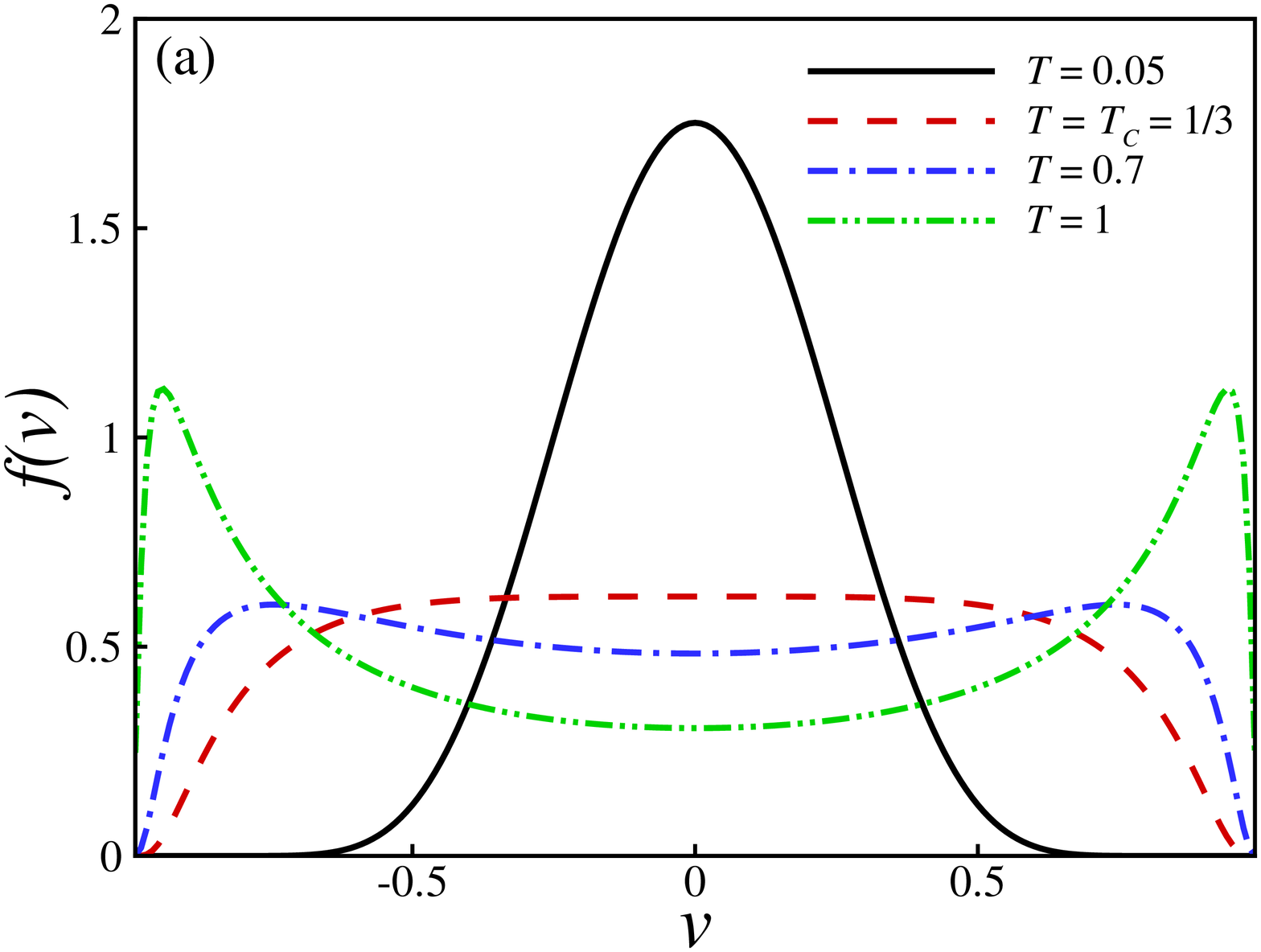,width=0.95\linewidth,clip=}\\
\epsfig{file=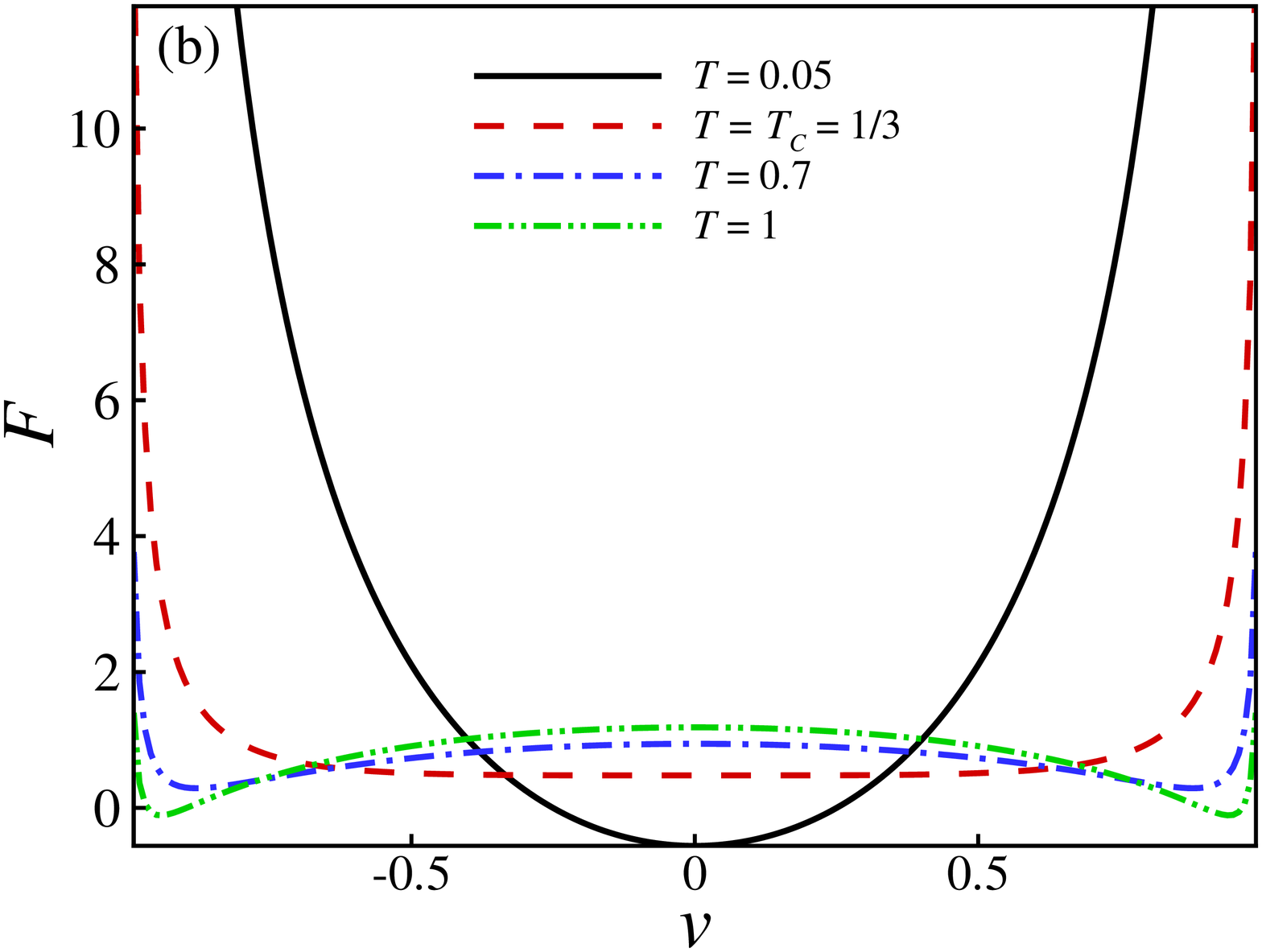,width=0.95\linewidth,clip=}
\\
\end{tabular}
\vspace*{0cm} \caption{(a) The J\"{u}ttner distribution function for $ d=1 $, for $ u=0 $, at temperatures $ T=0.05, 1/3, 0.7, 1 $ and (b) the corresponding thermodynamic function $F=-\ln f$.}\label{F1}
\end{figure}

As mentioned above, the phase transition under consideration is of a morphological nature as opposed to standard macroscopic transitions due to collective behavior of microscopic constituents. It is therefore not surprising that the order parameter, $v_{mp}$, is of a morphological nature, i.e. the peak of the probability distribution. Note that the average velocity is always zero in the co-moving frame $<\vec{v}>=\vec{u}=0$ regardless of temperature due to isotropic symmetry ($\vec{v}\rightarrow -\vec{v}$). Therefore, the transition is associated with emergence of non-zero $v_{mp}$ which signals departure from classical Gaussian results, thus indicating that $T_{c}$ \emph{is the scale} for the classical-relativistic crossover. On the other hand, $\chi$ measures how such an order parameter is susceptible to small changes in average velocity, $u$, near the critical point. Due to the flatness of the J\"{u}ttner distribution at the critical point (see Fig.1) even a small amount of  $u$ leads to a breaking of symmetry and a significantly large $v_{mp}$, and thus a diverging susceptibility. We note that it is the existence of diverging response function that is a true characteristic of a critical phase transition which was missing in the previous study \cite{mendoza}. Furthermore, the symmetry that is spontaneously broken is caused by velocity upper-bound constraint $(v\leq1)$ which limits arbitrary increases and therefore piles up a significant number of particles towards the upper limit as temperature rises.  Therefore, the symmetry breaking is associated with relativistic constraint, which does not break the isotropic symmetry ($u=0$) but leads to a spontaneous emergence of a non-zero peak in the velocity distribution.  We have therefore provided answers to all the questions posed in the Introduction.
\begin{figure}
\centering
\begin{tabular}{cc}
\epsfig{file=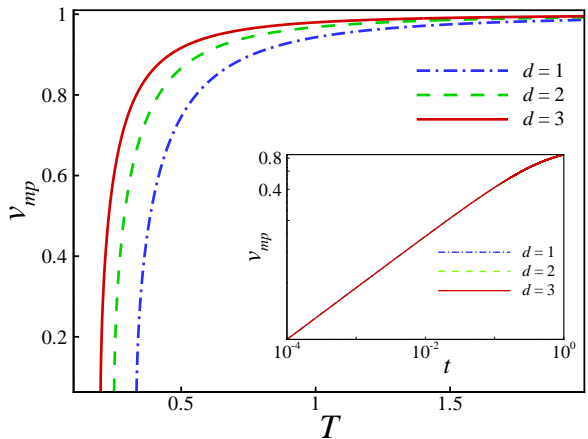,width=0.95\linewidth,clip=}
\end{tabular}
\vspace*{0cm} \caption{Temperature dependence of the order parameter, $v_{mp}$, in different dimension, for $ u=0 $. The inset shows double-logarithmic plot of the order parameter as a function of the rescaled temperature $ t=(T-T_{c})/T_{c}, $ confirming $v_{mp}\sim t^{\frac{1}{2}}$. The plots are obtained using numerically exact solution of Eq.(1) for $v_{mp}$. The curves for various dimensions in the inset fall on the top of each other, showing the independence of results from the physical dimension $d$. }\label{F2}
\end{figure}
\begin{figure}
\centering
\begin{tabular}{cc}
\epsfig{file=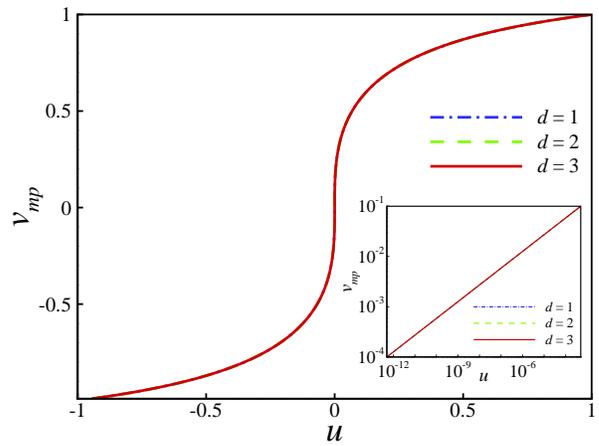,width=0.95\linewidth,clip=} \\
\end{tabular}
\vspace*{0cm} \caption{The order parameter versus the conjugate field at $T=T_{c}$ for various dimensions. The inset shows the corresponding double logarithmic plot confirming $v_{mp}\sim u^{\frac{1}{3}}$. The curves are obtained using numerically exact solution of Eq.(1) for $v_{mp}$. The curves for various dimensions fall on the top of each other, showing the independence of results from the physical dimension $d$. }\label{F3}
\end{figure}

A few comments are in order here. First, our thermodynamic potential is not a thermodynamic function in a proper sense, since it depends on microscopic quantities. It is important to note that we construct such a potential in analogy with free energy in order to be able to use the well-known formalism of a thermodynamic phase transition. Secondly, our thermodynamic potential is different from the standard Landau potential where $b(T)$ is taken to be positive for all $T$ as required by thermodynamic stability, because in our thermodynamic potential Eq.(3), $b(T)$ becomes zero well away from the transition point, i.e. at $T=(3/2)T_{c}$. In fact all higher order terms have this property that they have coefficients that become zero for increasingly larger temperatures away from the critical point, as shown below:
\begin{figure}
\centering
\begin{tabular}{cc}
\epsfig{file=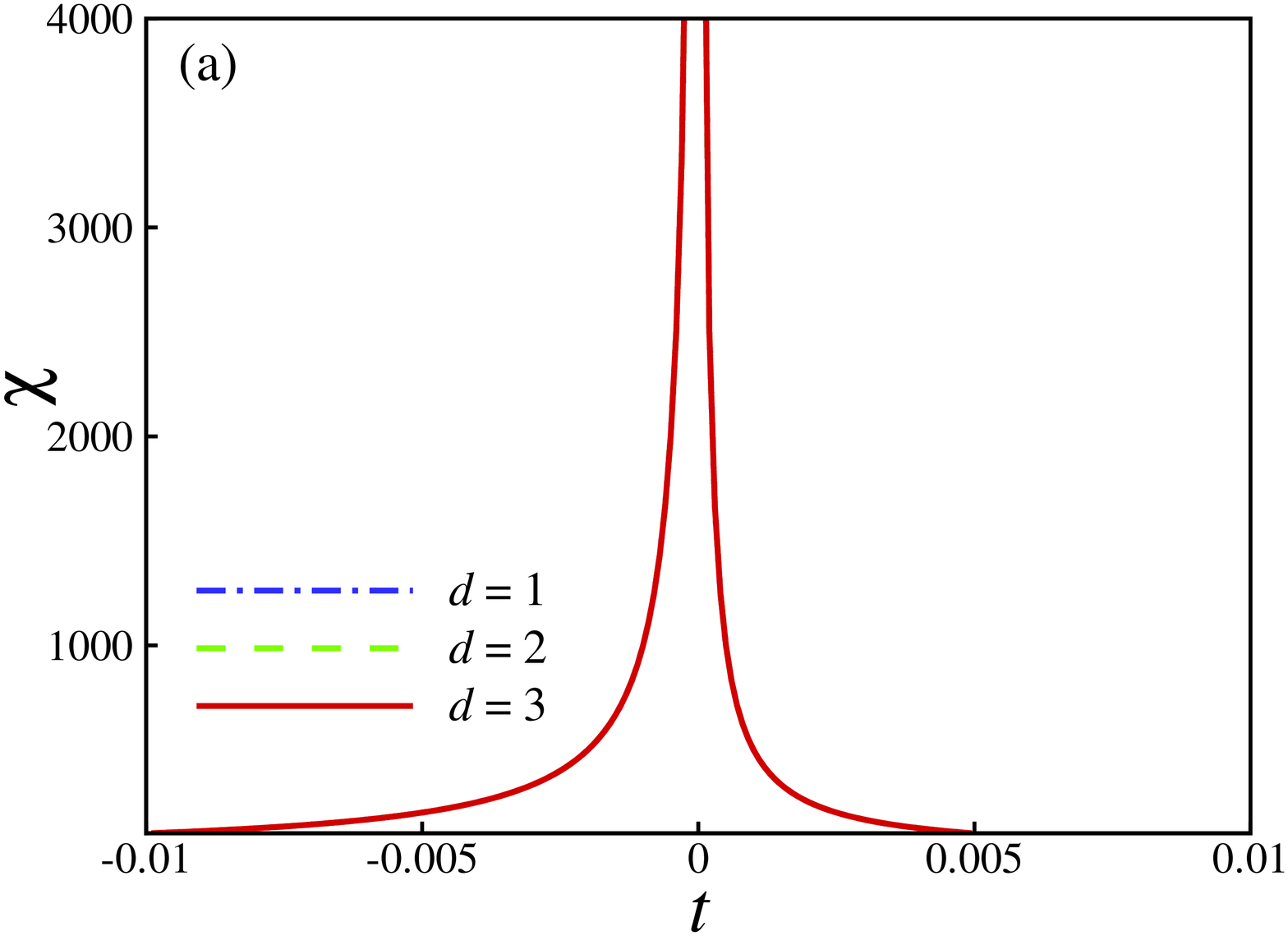,width=0.95\linewidth,clip=}\\
\epsfig{file=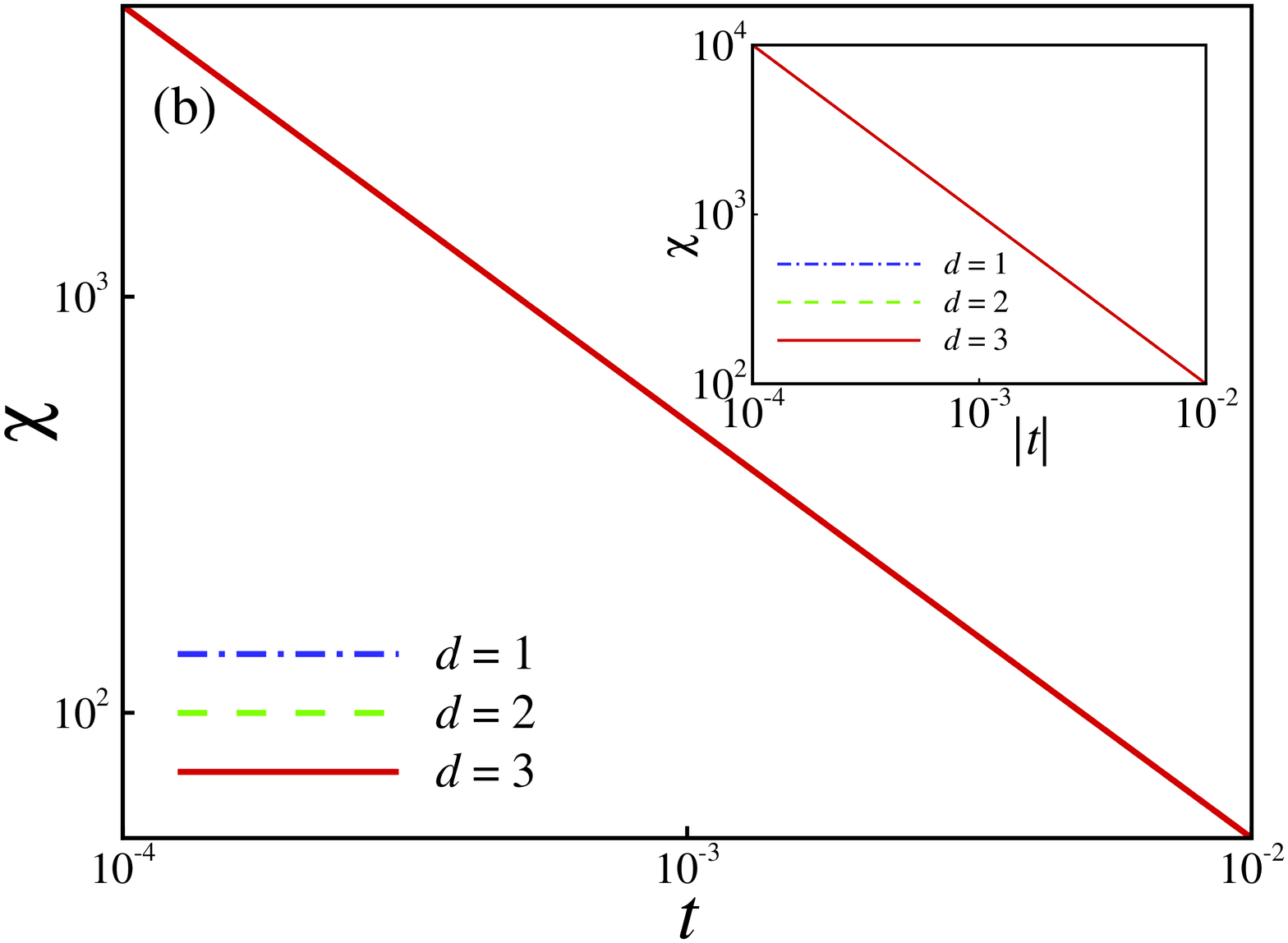,width=0.95\linewidth,clip=}
\end{tabular}
\vspace*{0cm} \caption{(a) The susceptibility as a function of the rescaled temperature near the critical temperature and (b) its double logarithmic plot confirming that $\chi\sim|t|^{-1}$ near criticality. The plots are obtained using numerically exact solution of Eq.(1) for $v_{mp}$. The inset in part (b) shows the behavior below transition. The curves for various dimensions fall on the top of each other, showing the independence of results from the physical dimension $d$. }\label{F4}
\end{figure}
\begin{eqnarray}
F(\vec{v},\vec{u},T)&=&\big[-\ln A+\frac{1}{T}\big]+\frac{1}{2}[\frac{1}{T}-\frac{1}{T_{c}}]v^{2} \nonumber\\
&+&\frac{3}{8}\big[\frac{1}{T}-\frac{2}{3T_{c}}\big]v^{4}+\frac{5}{16}\big[\frac{1}{T}-\frac{8}{15T_{c}}\big]v^{6} \nonumber\\
&+&\frac{35}{128}\big[\frac{1}{T}-\frac{16}{35T_{c}}\big]v^{8}-\frac{\vec{u}.\vec{v}}{T}+\ldots \
\end{eqnarray}

Note that the order parameter cannot be larger than unity and thus the problem of stability for arbitrary large order parameter is not relevant here, see also the comment following Eq.(6). Clearly, the concavity of our thermodynamic functions can be seen from Fig.1. However, in order to check the relations obtained from our expansion (Eqs.(7--9)) we have obtained numerically exact solutions to Eq.(1) and have plotted the corresponding relations in Figs.2--4. Fig.2 shows the emergence of the order parameter near the critical point in the absence of $u$ for various dimensions. Fig.3 shows the change of $v_{mp}$ as a function of $u$ at criticality, and Fig.4 shows the results for susceptibility both above and below the critical point.  These results confirm that $\beta=\frac{1}{2}$, $\delta=3$  and $ \gamma=1.$

Furthermore, in the theory of phase transitions there are typically six exponents which describe the thermodynamic behavior of systems at and near criticality. In addition to the three exponents reported above, there are $\alpha$, $\nu$ and $\eta$ \cite{goldenfeld}. We suspect that the heat capacity exponent $\alpha$ will not have much relevance here, but the other two exponents which have to do with fluctuations and correlations in order parameter may have important implications in relativistic systems. In the classical picture, the Markovian nature of the particle velocity in subsequent collisions leads to a Gaussian probability distribution for a random walk. However, in the relativistic systems, the upper-bound limit to velocity prohibits the Gaussian behavior where memory effects introduced into velocity correlation functions lead to non-Markovian processes \cite{Dunkl,pre,hakim}. Clearly, the role of such correlations must become more important as temperature rises, however, whether such correlations play a significant role and/or can be related to the critical behavior described here is an important and non-trivial question which are best addressed by a microscopic approaches such as molecular dynamics.

Finally, it is worthwhile to make a few comments about the inclusion of chemical potential ($\mu\neq0$) and/or the effect of quantum statistics ($\lambda\neq0$). It can easily be seen that non-zero values of $\mu$ for Boltzmann statistics does not change the singular behavior of our thermodynamic potential in Eq.(3). However, we find that the inclusion of quantum statistics ($\lambda\neq0$), while it does not change our general formalism, changes the value of $T_{c}$ in the deep quantum regime of $\mu\gtrsim0$. This modification is such that in the case of Bosons ($\lambda=-1$) $T_{c}$ is increased slightly as one approaches the quantum regime of $\mu\approx0$. But more interestingly, for the case of Fermions ($\lambda=1$), $T_{c}$ is reduced significantly as $\mu$ is increased beyond $\mu=0$. This reduction of $T_{c}$ in the case of Fermions is important as it makes the relativistic regime much more accessible with current laboratory limitations. A good example of such an application may be plasma physics or graphene as already discussed in \cite{mendoza}. Accordingly, we have plotted $T_{c}$ in Fig.5 as a function of $\mu$ for various statistics by obtaining numerical solutions to Eq.(1) for the particular case of $d=2$. Similar results are obtained for other values of $d$.
\begin{figure}
\centering
\begin{tabular}{cc}
\epsfig{file=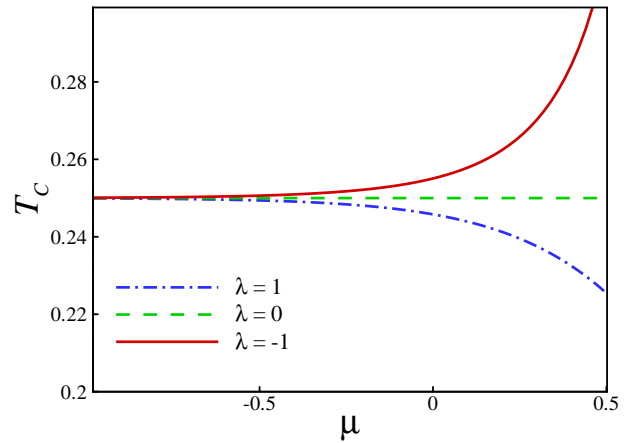,width=0.95\linewidth,clip=} \\
\end{tabular}
\vspace*{0cm} \caption{Critical temperature versus the chemical potential, for different statistics and $d=2$. These results are obtained by numerically exact solutions of Eq.(1)}\label{F5}
\end{figure}

\section{Concluding Remarks}
In this work we have used a general framework which connects statistical mechanics with thermodynamics, in order to study a recently proposed morphological transition in a relativistic gas from a thermodynamic point of view. We are able to obtain a Landau-like thermodynamic functional which immediately leads to critical transition at $T_{c}=1/(d+2)$ with mean-field critical exponents. While our results are consistent with the original study of such a transition \cite{mendoza}, our approach provides a general framework which not only leads to new results, but also helps in a better understanding of previously obtained results. Therefore, a relativistic gas is one whose most probable velocity is no longer zero, whereas in a classical (MB) gas $v_{mp}=0$ is always true regardless of temperature.  Even at $T=T_c$(i.e. at $k_BT=mc^2/5$ in three dimensions) where $v_{mp}$ is still zero, the velocity distribution has a broad (flat) range in contrast to classical Maxwellian distribution, see e.g. Fig.~1.  One would therefore expect that the kinetic behavior of the system would be different from the classical behavior.  One certain area which may provide some physical effects is the transport coefficients in a relativistic fluid which has recently gained some attention \cite{Kremer,Cercignani,Karlin}.  Particularly, the transport of heat due to particle gradient is negligible in a classical fluid, but becomes considerable in the relativistic limit \cite{pre,garcia}. Thus, the ratio of the transport coefficients which measures the relative share of the two mechanisms for transport (that due to particle gradient divided by that due to temperature gradient) is nearly zero in the classical limit ($T \ll 1$) and becomes approximately $1/3$ in the extreme relativistic regime ($T \gg 1$). Now, it is interesting to note that such a ratio reaches a significant and non-classical value of about $1/6$  around $T=T_c=1/5$, a considerable deviation from the classical results at a relatively low temperature. Another important quantity to look at might be the average energy per particle in the context of equipartition \cite{physrev}. Again, this function shows a smooth change for classical result of $dk_BT/2$ to extreme relativistic case of $dk_BT$ with significant change around $T_{c}$ (see Fig.1 in \cite{physrev}).

We therefore do not expect the leading thermodynamic functions of the system to exhibit critical (non-analytic) behavior at or around $T_{c}$, as would be expected in a real thermodynamic phase transition. However, it is clear that the thermodynamic behavior of a (highly) relativistic gas is distinctly different from its classical counterpart, and we expect that this change of behavior is dictated in a ``cross-over" region around $T_{c}$ where significant change in systems behavior occurs. We have therefore avoided the phrase ``phase transition" in this paper.

\end{document}